# CERN ACCELERATOR STRATEGY

S. Myers

*CERN, Geneva, Switzerland*

*Abstract*

The CERN strategy for future accelerator projects is outlined and the role of the HE-LHC inside this strategy is described.

## INTRODUCTION

The EuCARD-AccNet workshop HE-LHC'10 on a higher-energy LHC (HE-LHC) had invited a presentation on motivation, status, and strategy for HE-LHC studies. The motivation for the HE-LHC should come from the users, i.e. from the particle physicists, and it was already described by the previous speaker [1]. The present HE-LHC status covering magnets, detectors, cryogenics, vacuum, beam dynamics, injectors, etc. should come from the four main workshop sessions. The accelerator strategy, indeed, should come from the CERN Directorate. It is sketched in the following.

## STRATEGY

CERN has been, is, and will be the world's energy frontier laboratory. Presently, the LHC is being commissioned with beam. The LHC, with finally 7 TeV/beam, will be the highest energy collider on the planet for the foreseeable future. The higher-luminosity LHC (HL-LHC) is a proposed luminosity upgrade for installation in 2020–2021 and operation until around 2030. The HL-LHC also includes an upgrade of the LHC injector complex.

A study for an electron–proton collider based on the LHC, namely a Large Hadron electron Collider (LHeC) is supported by NuPECC and ECFA, and a Conceptual Design Report (CDR) is due to be finalized at the end of 2010 or early in 2011.

On the electron–positron front, the CLIC linear collider study will complete a CDR by 2011 and the CLIC Technical Design Report (TDR) by 2016–2020, depending on funding.

In the long-term strategic view of CERN, a Linear Collider would be constructed probably after the HL-LHC (>2030). BUT the question arises what will happen if the Linear Collider "does not fly" (e.g., for reasons of politics, finances, governance, energy and climate situation). What alternatives would exist in such a case? It seems there are two, namely HE-LHC and neutrinos. A project on the scale and innovation level of the HE-LHC has a long preparation lead time. Therefore, the HE-LHC'10 workshop appears timely. It complements the studies by a small HE-LHC working group which has been active at CERN since April 2010, and recently published its first considerations on the HE-LHC [2].

## SUMMARY

The CERN accelerator strategy comprises the following ingredients:

- LHC operation at 7 TeV/beam up to design luminosity;
- HL-LHC for installation in 2020/2021;
- Linear collider TDR for 2016–2020;
- Investigation of the HE-LHC as a feasibility study;
- R&D on high power proton drivers; and
- CDR for a LHeC (with ring–ring and ring–linac options).

## ACKNOWLEDGEMENTS

I thank the workshop organizers and the editors of the proceedings for their help in advancing the HE-LHC.